# A Geometric Origin for the Madelung Potential

D.H. Delphenich†
Physics Department
University of the Ozarks
Clarksville, AR 72840

*Abstract. Madelung's hydrodynamical forms of the Schrödinger equation and Klein-Gordon equation are presented. The physical nature of the quantum potential is explored. It is demonstrated that the geometrical origin of the quantum potential is in the scalar curvature of the metric that defines the kinetic energy density for an extended particle and that the quantization of circulation (Bohr-Sommerfeld condition) is a consequence of associating an* SO(2) *reduction of the Lorentz frame bundle with wave motion. The Madelung equations are then cast in basis-free form in terms of exterior differential forms in such a way that they represent the equations for a timelike solution to the conventional wave equation whose rest mass density function satisfies a differential equation of the "Klein-Gordon minus nonlinear term" type. The role of non-zero vorticity is briefly examined.*

**0. Introduction.** In 1927, the same year that Born, Heisenberg, and the others put forth the now-accepted statistical interpretation of the laws of quantum physics, Ernst Madelung [**1**] proposed a different interpretation that was based on a hydrodynamical interpretation. Although not widely accepted at the time, there was, nevertheless, a certain tributary of research that pursued this possibility. De Broglie [**2**] gave a more optical interpretation for the same basic equations. Takabayasi [**3,4**] expanded on the nature of the quantum corrections to the classical hydrodynamical equations, in both the non-relativistic and relativistic formulations. Vigier [**5**] made some progress toward incorporating general relativistic considerations. Various attempts were made to incorporate spin into the model at both the non-relativistic and relativistic levels [**6,7**].

Although, as we shall see, the equations of the Madelung model are essentially equivalent to those of the Schrödinger equation, the main reason that the Schrödinger formulation was accepted was that the statistical interpretation appealed to the intuition of the experimental physicists more than the "hydrodynamical" picture proposed by Madelung. Although Einstein always had grave reservations about the wisdom of leaving a random component in a theoretical model – an aspect he considered a sign of incompleteness – the statistical interpretation became the law of the land. Perhaps the main reason for this is that the main support for quantum theory came from its experimental verification. The statistical interpretation, with its reliance upon the measurement process and a statistical approach to the results of measurements, was very much the language and viewpoint of experimental physics. Furthermore, the vast majority of physicists are more experimentally than theoretically inclined, so there is a certain element of democracy to the acceptance of the statistical interpretation.

If there need be a justification for re-examining the Madelung model – beyond morbid curiosity – one should consider that there is a lot of mathematics and physics that has evolved since 1927, even though the basic "foundations" of quantum theory have changed very little since the days of the Copenhagen school, and some of the basic problems, such as the relationship between the foundations of quantum theory and the foundations of relativity theory, the question of pointlike or extended matter, and the true nature of quantization and renormalization, remain just as perplexing to this day. Perhaps taking a different viewpoint on the interpretation of the basic notions of quantum physics might illuminate some dark corner that the statistical interpretation leaves out.

---

† ddelphen@ozarks.edu

**1. The Madelung equations.** If we start with the Schrödinger equation:
$$\frac{\hbar}{i}\frac{\partial \Psi}{\partial t} + H\left(x, \frac{\hbar}{i}\nabla\right) = 0,$$

and express the complex wavefunction in polar form:
$$\Psi(t, x) = R e^{iS/\hbar},$$

then a straightforward calculation gives us a pair of equations that are obtained from the real and imaginary parts of the original equation, namely:
$$\frac{\partial S}{\partial t} + H(x, \nabla S) - \frac{\hbar^2}{2m}\frac{\Delta R}{R} = 0$$
$$\frac{\partial P}{\partial t} + \frac{\partial}{\partial x^i}(P\dot{x}^i) = 0 \qquad (\dot{x}^i \equiv \left[\frac{\partial H}{\partial p_i}\right]_{p=\nabla S}, \quad P \equiv R^2).$$

The first equation has a "Hamilton-Jacobi" sort of form ([1]) and the second has the appearance of a continuity equation for P if we identify $p = \nabla S$ with momentum.

If we assume that P, which in the context of the Schrödinger model represents the probability density function for the position of the particle that is described by the wave function $\Psi$, still represents the density of "something" in the present context, and **v** is the velocity vector field for the motion of whatever type of matter P describes then we seem to have a model for the motion of some sort of continuous medium. Indeed, if we make the usual expression for the H as $H(x,p) = \frac{p^2}{2m} + V(x)$ then the gradient of the Hamilton-Jacobi equation takes the form of an Euler equation:
$$m\frac{d\mathbf{v}}{dt} = -\nabla V + \frac{1}{P}\nabla\left(\frac{\hbar^2}{2m}\frac{\Delta R}{R}\right).$$

In this form we seem to be looking at a classical Newtonian equation of motion for the mass $m$ with an extra force term that seems to be of purely quantum origin, at least if the appearance of the factor $\hbar$ is any bellwether. The fact that the momentum field − hence the velocity field − is a gradient implies that the flow it describes is irrotational, and the continuity equation says that it is compressible as well. Evidently, understanding physical nature of the model hinges crucially upon understanding the nature of the quantum force term. A few more transformations of the form of this equation will help.

---
[1] The fact that R is coupled to the second equation makes this identification imprecise.

Since representing mass as a single number *m* suggests a point particle, and P is a density, which suggests an extended one, let us treat the constant *m* as a normalization constant and multiply the Euler equation through by P. We then identify the product *m*P as the mass density $\rho$ of an extended particle whose shape is dictated by P. This is reasonable since all we are saying is that the probability of finding matter is a volume of space should be proportional to the mass it contains. Indeed, this sounds more like an equivalence than an equation ([2]).

The Madelung equations now take the form:

$$\frac{\partial S}{\partial t} + H(x, \nabla S) - \frac{\hbar^2}{2m} \frac{\Delta \sqrt{\rho}}{\sqrt{\rho}} = 0$$

$$\frac{\partial \rho}{\partial t} + \frac{\partial}{\partial x^i}(\rho \dot{x}^i) = 0.$$

Observe that the quantum force $\nabla \frac{\Delta \sqrt{\rho}}{\sqrt{\rho}}$ vanishes iff $\sqrt{\rho}$ is an eigenfunction of the Laplacian, i.e., iff $\Delta \sqrt{\rho} = \lambda \sqrt{\rho}$ for some real number $\lambda$.

With the substitutions described above, and calling $v^i = \dot{x}^i$ we can now express our Euler equation in the more suggestive form:

$$\frac{\partial}{\partial t}(\rho v^i) + \frac{\partial}{\partial x^k}(\rho v^i v^k) = -\frac{\rho}{m}\frac{\partial V}{\partial x^i} + \frac{\partial}{\partial x^k}\tau^{ik}.$$

Following Takabayasi [**3**, **4**], we have re-expressed the quantum force term as the divergence of a "quantum stress" tensor:

$$\tau_{ij} = \left(\frac{\hbar}{2m}\right)^2 \rho \frac{\partial^2 (\ln \rho)}{\partial x^i \partial x^j}.$$

To get a better picture of what sort of medium we are dealing with, we examine the character of the stress tensor. Since $\tau_{ij}$ is symmetric, we put it into diagonal form by changing to normal coordinates. The diagonal elements, when expressed in normal coordinates – i.e., the *principal stresses* – take the form:

$$p_i = \frac{\hbar^2}{2m^2} \rho \frac{\partial^2 \sigma}{\partial (x^i)^2},$$

with $\rho = e^{2\sigma}$. $p_i$ is called *tension-like* if $p_i > 0$, and *pressure-like* if $p_i < 0$. The *mean pressure* is given by:

---

[2] Certainly, it says nothing about massless matter, but for massless particles one needs to use relativistic methods anyway; we shall discuss these methods in due course.

$$\bar{p} = -\frac{1}{3}\text{Tr}(\tau_{ij}) = -\frac{\hbar^2}{6m^2}\rho\Delta\sigma.$$

It should be pointed out that this pressure can be negative, as well as positive. In classical hydrodynamics [**8**] negative pressures are often associated with cavitation, which involves the formation of topological defects in the form of bubbles.

In order for the medium in question to be an ideal fluid, we would have to have:
$$\tau_{ij} = -\bar{p}\delta_{ij}.$$

This case occurs when and only when the mass density is Gaussian:
$$\sigma \propto -x^i x_i,$$
in which case:
$$\bar{p} \propto \frac{\hbar^2}{2m^2}\rho.$$

Part of the transition from quantum mechanics to classical mechanics at this step takes the form of the transition from an extended particle to a point particle, in which the Gaussian distribution goes to the Dirac distribution.

Since the aforementioned stress tensor is not generally isotropic, apparently our Madelung medium is not necessarily an ideal fluid. Moreover, if it were a viscous fluid, we should expect to see $\tau_{ij}$ coupled to the rate of deformation tensor, which is derived from $D\mathbf{v}$. Since this does not seem to be the case, rather than commit ourselves to calling this form of matter a fluid medium, we refer to a space given a mass density function $\rho$, a velocity vector field $\mathbf{v}$, and a stress tensor $\tau_{ij}$, defined as above, as a *Madelung continuum.* This would represent something like an inviscid fluid that also supports shear stresses, whereas the Gaussian wave packet of quantum mechanics corresponds to an ideal compressible irrotational fluid medium.

It is not unreasonable that the continuum mechanics of elementary matter should involve a state of matter that is not identical with the known macroscopic forms, when one remembers that the forms of matter that are treated by conventional continuum mechanics, such as solids and fluids, are always assumed to be reducible to "molar" ensembles of a large number of more elementary constituents, and their material properties are derived from the type of interactions that one assumes exist between the constituents.

If we add time as the zeroth coordinate, and extend the velocity vector by $v^0 = 1$, a process which could be called a *Galilean embedding* of our Newtonian problem in $\mathbb{R}^4$, then by defining the energy-momentum tensor as:

$$\mathcal{T}^{\mu\nu} \equiv \rho\left[v^\mu v^\nu - \left(\frac{\hbar}{2m}\right)^2 \frac{\partial^2(\ln\rho)}{\partial x^\mu \partial x^\nu}\right],$$

where we have extended the quantum stress tensor with zeros when the index is 0, we find that the pair of Madelung equations – Euler and continuity – can be consolidated into one equation([3]):

$$\frac{\partial \mathcal{T}^{\mu\nu}}{\partial x^\mu} = -\frac{\rho}{m}\partial^\nu V.$$

Note also that the classical potential gradient gets multiplied by the term $\rho/m = P$ so there seems to be a "quantum perturbation" to the classical potential in which some of the potential energy is transferred to the stress tensor. This seems consistent with the usual quantum thinking that measuring the state of a system changes the state of the system. Here the applied force not only accelerates the extended particle, but deforms it as well.

**2. The relativistic form of the Madelung equations.** To understand the geometric origin of the quantum stresses, it helps to look at the relativistic form of the Madelung equations. Fortunately, the basic nature of the equations does not change, although some mathematical details become significant.

First, we start with the (free massive spinless complex scalar particle) Klein-Gordon equation ([4]):

$$\left\{-\hbar^2 \eta^{\mu\nu}\frac{\partial^2}{\partial x^\mu \partial x^\nu} + m_0^2 c^2\right\}\Psi = 0.$$

The same sort of polar substitution for $\Psi$ as before gives a pair of equations for the real and imaginary parts:

$$\eta^{\mu\nu}\frac{\partial S}{\partial x^\mu}\frac{\partial S}{\partial x^\nu} + m_0^2 c^2 - \hbar^2\frac{\Box R}{R} = 0$$

$$\eta^{\mu\nu}\frac{\partial}{\partial x^\mu}\left(P\frac{\partial S}{\partial x^\nu}\right) = 0.$$

By defining four-velocity, rest mass density, energy-momentum, and the stress tensor:

$$u_\mu = \frac{1}{m_0}\frac{\partial S}{\partial x^\mu}, \quad \rho = m_0 P, \quad p_\mu = \rho u_\mu, \quad \tau_{\mu\nu} = \left(\frac{\hbar}{2m_0}\right)^2\frac{\partial^2(\ln\rho)}{\partial x^\mu \partial x^\nu},$$

and the energy-momentum tensor:

$$\mathcal{T}_{\mu\nu} = \rho[u_\mu u_\nu + \tau_{\mu\nu}],$$

we arrive at a concise form for the relativistic equations of motion for the medium described by $\rho$ and $u_\mu$:

---

[3] Admittedly, this is a somewhat misleading form, since we are still just talking about non-relativistic mechanics, but it still sets the stage for the relativistic form.

[4] Just to set the sign convention, we agree that $\eta^{00} = 1$ and $\eta^{ii} = -1$ for $i = 1, 2, 3$, and all other components are equal to 0.

$$\partial^\mu \mathcal{T}_{\mu\nu} = 0, \qquad\qquad \partial^\mu p_\mu = 0.$$

The first equation is a relativistic form of the previous non-relativistic equation in the absence of external forces. The second equation represents a relativistic statement of incompressibility [**9,10**], which does not contradict the non-relativistic compressibility of the medium since – in the eyes of relativity – incompressibility in fluid media is equivalent to an infinite speed of light, as is rigidity in solid media.

Our present task is to recast the relativistic form of the Madelung equations in a form that exhibits the interplay between geometry and continuum mechanics as it relates to wave motion.

**3. The representation of quantum wavefunctions by spacelike 2-frames.** The support of a quantum (i.e., Schrödinger or Klein-Gordon) wave function $\psi = \mathrm{R}e^{iS/\hbar}$ takes its values in the one (complex) dimensional abelian Lie group $\mathbf{C}^*$ of non-zero complex numbers under multiplication. The general element of this group takes the form: $z = e^{\sigma + i\theta}$. We have the isomorphism: $\mathbf{C}^* \cong \mathbb{R}^+ \times U(1)$, which shows that, up to homotopy, $\mathbf{C}^*$ is simply $S^1$.

Since the complex plane may also be represented by $\mathbb{R}^2$ as a point set, we can also represent $\mathbf{C}^*$ by linear transformations of $GL(2,\mathbb{R})$. This is effected by the (real) linear map of the Lie algebra of $\mathbf{C}^*$ into $\mathfrak{gl}(2,\mathbb{R})$ defined by $\sigma + i\theta \mapsto \sigma I + \theta J$, where I is the two-dimensional identity matrix and $J = \begin{bmatrix} 0 & -1 \\ 1 & 0 \end{bmatrix}$ defines a complex structure on $\mathbb{R}^2$. Exponentiation of the Lie algebras defines the corresponding map of the Lie groups. In particular, $e^{\sigma + i\theta}$ goes to $e^\sigma R_\theta$, where $R_\theta$ is the rotation matrix, $R_\theta = \begin{bmatrix} \cos\theta & -\sin\theta \\ \sin\theta & \cos\theta \end{bmatrix}.$

In order to justify that the quantum wavefunction has a (real) geometric origin, it would be reassuring to find a representation for $\mathbf{C}^*$ in terms of something that lives on spacetime in a natural way, either in its tangent bundle or its bundle of linear frames. The approach we shall take is based in the notion, which is discussed in more detail in [**11**], that wave motion in general defines a pair of transverse foliations of codimension one of the region of spacetime in which the motion occurs.

The first foliation is a proper time foliation of spacetime by *simultaneity hypersurfaces* that are orthogonal to a timelike unit vector field **t**. The choice of this vector field is equivalent to a choice of rest frame for the motion to be described in. Dually, one could define the timelike 1-form $\theta = i_\mathbf{t} g$ and obtain the proper time foliation by integrating the sub-bundle of T(M) that is defined by the annihilating subspaces of $\theta$; of course, this assumes that the sub-bundle in question is integrable. The condition for integrability of the sub-bundle that is defined by $\theta$ is that its Frobenius form $\theta \wedge d\theta$ vanish.

The second foliation defined by a spacelike phase 1-form $\phi$; hence, the timelike leaves of this foliation are called *isophases*. Furthermore, we assume that the leaves of this foliation intersect the simultaneity leaves transversally ([5]). By using the Lorentz structure that is defined by **t** and a choice of spacelike metric to project $\phi$ onto a 1-form $\phi_\Sigma$ tangent to the simultaneity leaves, we can define a foliation of each simultaneity leaf by *momentary wave fronts*. The vector field that is metric-dual to $\phi_\Sigma$, when normalized, will be denoted by **n**.

Since any physical wave has to have a source, we shall have to deal with singular foliations where the singularities account for the wave sources. Hence, it will not be necessary to assume that the foliations are global, or, as a consequence, that the topological obstructions to their global existence vanish. Physically speaking, one generally thinks of matter as having something of a spatially localized nature, which is consistent with the nature of singularities in their least pathological cases.

The pair of orthogonal vector fields, **t** and **n**, define a (possibly singular) orthonormal 2-frame field on M. Hence, they also define a reduction of the bundle of Lorentz frames wherever both vectorfields non-zero, i.e., a *G-structure* on M-{singularity subset} where G = SO(2). All that one has to do is to restrict oneself to Lorentz 4-frames that have **t** and **n** as members. The only "gauge" freedom left in the definition of the frame field is the two-dimensional plane orthogonal to the span of **t** and **n**. This means the structure group has been reduced from GL(4) to SO(3) to SO(2). In particular, the reduction from GL(4) to SO(3) was defined by the vector field **t** and the second reduction from SO(3) to SO(2) was defined by **n**. If we were expecting **t** and **n** to define a non-singular orthonormal 2-frame then we would be assuming that spacetime has degree of parallelizability at least two. This would then entail the vanishing of the top two Stiefel-Whitney classes of GL(M), $w^3(M)$ and $w^4(M)$.

In order to account for the fact that our quantum wavefunction has a scaling factor that varies as a function of time and spatial position − namely $\rho$ − we point out that reductions of GL(M) to G-principal sub-bundles are not general unique. For instance, since any given linear frame in a particular tangent space defines a corresponding orbit under the action of G, unless there is some way to distinguish the orbit that one chooses at each point to define the fiber of the G-structure, one must treat any such choice as essentially equivalent. For instance, the orbits of the action of SO(2) on the plane are circles of radius $r$. In order to specify which orbit in $GL_x(M)$ was chosen to define the SO(2) reduction, one also needs to specify the radius of the circle. Hence, $\rho$ represents a choice of scale at each point.

We now represent the quantum wavefunction $\psi$ by a generic ([6]) section $h^i$ of this reduced bundle, SO(2)(M). We can represent $h^i$ as $h^i = h^i_j dx^j$ relative to the natural coframe field $dx^i$ that is defined by an adapted local coordinate system $x^i$, for which the plane orthogonal to **t** and **n** is the *xy*-plane, and **n** is a unit vector field in the *z*-direction. The matrix $h^i_j$ then takes the form:

---

[5] This condition actually defines the isophase foliation as a "foliated cobordism" between the foliations of the simultaneity leaves by momentary wavefronts. For more discussion of this, cf. [**11**].

[6] By *generic*, we mean that it is defined on an open subset of M whose complement is its singularity cycle.

$$h^i_j = \sqrt{\rho}\begin{bmatrix} 1 & 0 & 0 & 0 \\ 0 & \cos\alpha & -\sin\alpha & 0 \\ 0 & \sin\alpha & \cos\alpha & 0 \\ 0 & 0 & 0 & 1 \end{bmatrix},$$

Since the action of $\mathbf{C}^*$ on the Lorentz 4-frames is non-trivial on only two members, we shall henceforth think of $h^i$ as a spacelike orthogonal 2-frame in the plane orthogonal to the plane of **t** and **n** with a norm of $\sqrt{\rho} = $ R and an angular orientation of $\alpha = S/\hbar$. (Hence, $i$ and $j$ will range over 1 and 2.)

We shall now examine the geometric and topological consequences of our decision to represent a quantum wave function by pair consisting of an orthonormal 2-frame field and a scaling factor. In particular, we shall examine the fact that such a pair, or, equivalently $\{\rho, \alpha\}$ essentially amounts to a choice of "gauge" for our wave structure.

**4. The angular gauge structure.** For a given choice of $\rho$, our 2-frame field $h^i$ becomes orthonormal if we regard the factor $\sqrt{\rho}$ as rescaling our choice of unit norm. In the last section, we defined the angular orientation of this 2-frame with reference to the coordinate frame $\{dx, dy\}$, but, in general, the description of the angular orientation of $h^i$ by an angle at each point is only defined up to an arbitrary choice of zero angle. Hence, to choose $h^i$ is to choose an SO(2) gauge for our wave structure $\{\theta, \phi\}$, so we need to examine what is involved with making our description of wave motion independent of this choice.

If $\bar{h}^i$ is another choice of gauge whose domain of definition overlaps that of $h^i$ then $\bar{h}^i$ will be related to $h^i$ by a transition function $R_\alpha$ that is defined on the overlap and takes its values in SO(2), which is represented by 2×2 real rotation matrices that are defined with respect to $h^i$. Hence:

$$\bar{h}^i = R_\alpha h^i.$$

When one transforms the differential of $h^i$, one defines a 1-form $\varpi$ with values in the Lie algebra of SO(2), which is the imaginary line:

$$D\bar{h}^i = D(R_\alpha h^i) = DR_\alpha \otimes h^i + R_\alpha Dh^i = R_\alpha(Dh^i + \varpi \otimes h^i),$$

in which we have defined:

$$\varpi = R_\alpha^{-1} DR_\alpha = \varphi J, \qquad \varphi = d(\ln \alpha).$$

The 1-form $\varphi$ represents the "angular velocity" of the frame $\bar{h}^i$ with respect to $h^i$ at each point $x \in M$. It is also the term that gets added to the local representative of a more general SO(2) connection $\omega$ on SO(2)(M) when it gets transformed from $h^i$ to $\bar{h}^i$:

$$\omega \mapsto \omega + \varpi.$$

If we let $\omega = \chi J$ and $\varpi = d\lambda J$ then this gives rise to the usual form of an Abelian gauge transformation of the second kind:

$$\chi \mapsto \chi + d\lambda.$$

By a previous identification:

$$S = \hbar \alpha.$$

Hence, by another momentum takes the form:

$$p = dS = \hbar d\alpha = \hbar \alpha d\varphi.$$

We then generalize our definition of the momentum 1-form to be related to a choice of connection $\omega$ on SO(2)(M) by:

$$pJ = \hbar \omega.$$

The basic holonomy integral of $\omega$ around a loop $\gamma$, $\oint_\gamma \omega$, then gives us the *dynamical circulation* of $p$ around $\gamma$:

$$\Gamma[\gamma] = \oint_\gamma p = \hbar \oint_\gamma \omega.$$

Since $\gamma$ is homotopic to $S^1$ and the function $\int_{\gamma(0)}^{\gamma(\tau)} \omega$ takes its values in $S^1$ the integral must equal $2\pi n$, where $n$ is the winding number of the map from $\gamma$ to $S^1$. This means that circulation gets quantized according to the Bohr-Sommerfeld rule ([7]):

$$\Gamma[\gamma] = \oint_\gamma p = hn.$$

If we continue to regard momentum as proportional to an SO(2) connection 1-form $\omega$ then the 2-form $dp$, which has the interpretation of *dynamical vorticity* in hydrodynamics, is proportional to the curvature of that connection. If we go back to the fact that our plane of rotation is the plane orthogonal to the one spanned by **t** and **n**, hence, the plane tangent to the momentary wavefronts, then we see that $dp$ is proportional to the Gaussian curvature of these wavefronts.

By Gauss-Bonnet [**12**], the integral of $dp$ over a momentary wavefront will vanish as long as its Euler-Poincaré characteristic does, too. However, the curvature integral in Gauss-Bonnet is proportional to the first Chern number of the restriction of the bundle SO(2)(M) to the momentary wavefront, such this is also a statement about the triviality of that bundle: if the integral does not vanish then the bundle in question is not trivial and $\omega$, hence $p$, cannot be represented by a global 1-form. As for the second Chern class, since it is proportional to the Euler-Poincaré characteristic of M, which we assumed to vanish by defining a Lorentz structure, it will vanish as well.

We summarize this section by the statements:
    a)    The quantum wave function $\psi$ corresponds to a local section $h^i$ of SO(2)(M).
    b)    The momentum $p$ (= $dS$), is proportional to an SO(2) connection form $\omega$ on the

---

[7] This relationship also plays a fundamental role in the theory of quantum vortices [**13-16**].

principal bundle SO(2)(M).

c) The quantization of circulation for *p,* i.e., Bohr-Sommerfeld quantization rule, arises from the fact that $\pi_1(S^1) = \mathbb{Z}$.

d) The curvature of $\omega$ relates to the dynamical vorticity of *p* and the curvature of the momentary wavefronts.

**5. Geometric origin of the quantum potential.** Now, we examine the consequences of choosing a scale of unit norm for our frame by way of the function $\sqrt{\rho}$. We shall see that this decision is at the root of our inclusion of the quantum potential in our Madelung equations.

In general, when a manifold M is given a metric (or pseudo-metric) *g,* another metric $\bar{g}$ is said to be *conformally related* to *g* if there is a positive function $\Omega^2 > 0$ such that $\bar{g} = \Omega^2 g$. If we express $\Omega$ in the form $\Omega = e^\sigma$ then the way that the Levi-Civita connection, Ricci curvature, and scalar curvature change under a conformal change of metric are ([8]) [*cf.,* **17,18**]:

$$\bar{\Gamma}^i_{jk} = \Gamma^i_{jk} + \delta^i_j \partial_k \sigma + \delta^i_k \partial_j \sigma - g_{jk} g^{il} \partial_l \sigma$$

$$\bar{R}_{ij} = R_{ij} - (n-2)\sigma_{ij} - [\Delta\sigma + (n-2)(\partial^k \sigma \partial_k \sigma)]g_{ij}$$

$$\bar{R} = e^{-2\sigma}[R - 2(n-1)\Delta\sigma - (n-1)(n-2)(\partial^i \sigma \partial_i \sigma)].$$

We have defined the matrix $\sigma_{ij}$ as

$$\sigma_{ij} = \partial_i \partial_j \sigma - \partial_j \sigma \partial_i \sigma.$$

If you consider the difference between the spacetime metric *g,* which is defined on the tangent bundle here, and defines the differential of arc-length, or proper-time interval, of a curve segment, and the expression *mg* which defines (twice) the differential of kinetic energy of a *point-like* particle of mass *m* moving along that curve then there is little to distinguish between the two geometrically. This would represent a rather trivial sort of conformal change of metric.

Here is where the difference between point-like matter and extended matter becomes pronounced. If the constant *m* is replaced by the non-negative non-constant smooth function $\rho$ then one must contend with how its derivatives $\partial_i \rho$ affect the geometry of the new metric when they are non-vanishing.

In particular, let us look at how $\rho = e^{2\sigma}$ deforms the geometry of the Minkowski metric on $\mathbb{R}^4$. This means we are starting with $g = \eta$, $\Gamma^i_{jk} = 0$, $R_{ij} = 0$, $R = 0$, $n = 4$. The expressions become:

$$\bar{\Gamma}^i_{jk} = \delta^i_j \partial_k \sigma + \delta^i_k \partial_j \sigma - \eta_{jk} \eta^{il} \partial_l \sigma$$

$$\bar{R}_{ij} = -2\sigma_{ij} - [\Box\sigma + 2(\partial^k \sigma \partial_k \sigma)]\eta_{ij}$$

---

[8] Here, *n* represents the dimension of M.

$$\bar{R} = -6e^{-2\sigma}[\Box\sigma + (\partial^i\sigma\partial_i\sigma)].$$

If we substitute $\Omega$ back into the equation for scalar curvature, we get:

$$\bar{R} = -\frac{6}{\Omega^2}\left(\frac{\Box\Omega}{\Omega}\right).$$

This allows us to express our quantum potential in the form:

$$\frac{\Box\sqrt{\rho}}{\sqrt{\rho}} = \frac{\Box\Omega}{\Omega} = -\frac{1}{6}\bar{R}\Omega^2 = -\frac{1}{6}\rho\bar{R}.$$

What makes this result somewhat startling is that it says that the (mathematical) meaning of the quantum potential, which distinguishes Newton's from Schrödinger's equation of motion, is that we are coupling the scalar curvature times the mass density to the total energy of the system as a "quantum correction". This is entirely reasonable from the standpoint of general relativity, where a related expression describes the integrand in the Einstein-Hilbert action functional.

We can also observe that, from general geometric principles:

$$\partial^i\bar{R}_{ij} = \partial^i(\frac{1}{2}g_{ij}\bar{R}) = \frac{1}{2}\partial^j\bar{R},$$

so we can also say that since this implies that

$$\partial^i\bar{R}_{ij} = -3\partial^i\tau_{ij},$$

apparently the Takabayasi stress tensor differs from the Ricci curvature of the kinetic energy metric only by only a term which has vanishing divergence. This means that dynamically there is no loss of generality in using the Ricci curvature of $g = \rho\eta$ as the stress tensor, instead of the Takabayasi expression, since both define the same quantum force field. Otherwise stated: the quantum force is proportional to either the divergence of the Ricci curvature tensor of $g$ or the gradient of its scalar curvature. This also means we are dealing with principal curvatures instead of principal stresses.

To extend these considerations into the general relativistic framework, we first observe that all we have to do is use a more general Lorentz manifold than Minkowski space. That means starting with a $g_{ij}$ of Lorentz type, and a Levi-Civita connection and curvature that do not vanish. Under a conformal change from the spacetime metric to the energy metric, the Einstein tensor becomes:

$$\bar{G}_{\mu\nu} = \bar{R}_{\mu\nu} - \frac{1}{2}\bar{g}_{\mu\nu}\bar{R} = G_{\mu\nu} - 2\sigma_{\mu\nu} + [2\Box\sigma + \partial^\lambda\sigma\partial_\lambda\sigma]g_{\mu\nu}.$$

It seems reasonable to assume that this implies a "quantum correction" to the Einstein equation:

$$G_{\mu\nu} + (\Box\sigma + \frac{1}{2}\partial^\lambda\sigma\partial_\lambda\sigma)g_{\mu\nu} = 8\pi GT_{\mu\nu} + 2\sigma_{\mu\nu}.$$

We can also think in terms of a quantum correction to the equation of geodesic motion:

$$\nabla_v p = -[(\delta^i_j\partial_k\sigma + \delta^i_k\partial_j\sigma - g_{jk}g^{il}\partial_l\sigma)v^j p_i]\,dx^k.$$

$$= -(p(\mathbf{v})d\sigma + \mathbf{v}\sigma p - \mathbf{p}\sigma v)$$

(We have introduced the vector field **p** that is metric-dual to the 1-form $p$ and the 1-form $v$ that corresponds to the vector field **v**.) Such an equation can be interpreted variously as quantum fluctuations to the classical extremal or Newton's law of motion with a quantum force term. In the event that momentum is related to covelocity by the simple prescription $p = \rho_0 v$, and we recall that $\rho_0 = e^{2\sigma}$, the geodesic equation becomes simply:

$$\nabla_\mathbf{v} p = -\tfrac{1}{2} c^2 \, d\rho_0.$$

Once again, we see the geometric role of extended matter since the force that produces the quantum correction to the classical extremal originates in the difference between the particle in question, which is described by a non-constant rest mass density, and a pointlike one.

The deeper consequences of such modifications are beyond the scope of the present paper and will have to define a direction for future exploration. At this point, we can, nonetheless, observe that we seem to be including the factor:

$$\Box\sigma + \frac{1}{2} g(\nabla\sigma, \nabla\sigma),$$

as a sort of "cosmological constant," and the tensor $\sigma_{\mu\nu}$ as a quantum correction to the energy-momentum tensor which seems to originate in the process of metric deformation.

**5. Generalized formulation of Madelung equations.** The Madelung equations can also be given a "basis-free" form in terms of exterior differential forms:

$$\begin{cases} p^2 = -m_0^2 c^2 + \hbar^2 \dfrac{\Box\sqrt{\rho_0}}{\sqrt{\rho_0}} \\ \delta p = 0 \\ \quad p = dS. \end{cases}$$

Some comments must made at this point:

1) The first equation makes sense physically, since the right-hand side goes to $-m_0^2 c^2$, the rest mass energy($/c^2$) of the point particle, as $h \to 0$. This suggests that the quantum potential is a geometric correction to the rest energy of the particle associated with the fact that it is not precisely pointlike. However, we can also rearrange it as follows:

$$p^2 = \hbar^2 \left[ \frac{\Box\sqrt{\rho_0}}{\sqrt{\rho_0}} - k_C^2 \right] = (\hbar\kappa)^2.$$

In light of our earlier section on the geometric origin of the quantum potential, we rewrite this as:

$$p^2 = \hbar^2 \left[ -\frac{1}{6}\rho_0 \bar{R} - k_C^2 \right], \quad \text{i.e.,} \quad \omega_0^2 = -\frac{1}{6}\rho_0 c^2 \bar{R}.$$

As long as the scalar curvature is negative, this is consistent with the physical interpretation that instead of a classical point particle ($\rho_0 = m_0 \delta(\gamma)$), we have "deformed" the point into a "smeared" point and the quantum correction that must be made to the rest energy of the particle is the inclusion of the energy of deformation, which is proportional to the scalar curvature of the deformed metric.

We then see that we are defining the frequency-wavenumber 1-form $\kappa = \omega_0/c\, d\tau + k_i dx^i$ that is associated with the energy-momentum $p$, if we set:

$$\frac{\omega_0}{c} = \left( -\frac{1}{6}\rho_0 \bar{R} \right)^{1/2} \text{ and } k^2 = k_C^2.$$

2) The last of the three equations suggests two possible generalizations of the model: $p$ closed, but not exact, and $p$ not closed. If $p$ is to be interpreted as the momentum 1-form for an extended massive particle – which is described by $(\rho_0, u)$ – then $dp$ gets the interpretation of the *dynamical vorticity* of the flow defined by $p$.

3) In the event that we choose $p$ to be closed (or exact, for that matter), the last two equations say that $p$ is closed and co-closed, i.e.

$$dp = 0, \qquad \delta p = 0,$$

hence, a harmonic field (in the sense of [**19**]). In the relativistic case, this points to a wave-like nature for the momentum 1-form of $(\rho_0, u)$.

4) If the continuum-mechanical interpretation is truly fundamental, and not merely an amusing curiosity, then one must also solve the problem defined by the Correspondence Principle: show how, by starting with the Madelung equations, one can *deduce* the Schrödinger-Klein-Gordon equation as a consequence – preferably in a way that makes scientifically intuitive sense.

Let us now pursue the consequences of the fact that one usually expects that $p^2 = -\rho_0^2 c^2$; if we revert back to our earlier notation, $\rho_0 = m_0 R^2$, this implies that:

$$\Box R - k_C^2 R[1 - R^4] = 0.$$

This has a reasonable "Klein-Gordon minus nonlinear term" appearance that suggests that we have also deformed our quantum plane wave into something more physically reasonable. (Recall that the plane wave is just as physically fictitious as the delta function.). Although $0 \leq R \leq 1$, which makes one expect the $R^4$ term to be a negligible contribution to the linear one, if one imagines that R is a "smoothed" delta function then presumably the nonlinear term becomes significant in a small neighborhood of the center of the distribution.

We can now rewrite the fundamental system of equations in the physically reasonable form ([9]):

$$\begin{cases} p = dS \\ \delta p = 0 \\ p^2 = -\rho_0^2 c^2 \\ \Box R - k_C^2 R[1 - R^4] = 0 \quad (\rho_0 = m_0 R^2). \end{cases}$$

If we generalize $p$ to be closed, but not exact, which is reasonable, since we have defined S to be proportional to the angle function, then the equations take the form:

$$\begin{cases} dp = 0 \\ \delta p = 0 \\ p^2 = -\rho_0^2 c^2 \\ \Box R - k_C^2 R[1 - R^4] = 0 \quad (\rho_0 = m_0 R^2). \end{cases}$$

The first two equations make $p$ a harmonic field, or, if you prefer, an irrotational incompressible flow. The third one is the causality constraint that derives from the fact that the particle is massive. The last equation determines the shape of the rest mass density. This basically says that the rest mass density for a classical massive point particle has to get deformed from a delta function into an extended massive particle ($\rho_0$, $u$) with a rest mass density that satisfies the last equation, so that $p$ becomes basically a (non-characteristic) solution of the wave equation. Presumably, for other examples of quantum systems, the main difference should be in the last equation, which seems to dictate the equilibrium shape of the particle. Hence, we have transformed the search for a geometrical interpretation of wave mechanics into the search for physical first principles that produce the last equation as a natural consequence; this undoubtedly involves reaching a deeper understanding of the role of geometry and topology in continuum mechanics.

**6. The introduction of nonzero vorticity.** Another issue that can be raised by the Madelung equations is what would be involved with dropping the assumption that $p$ be closed. In the preceding section, we considered the case of $p$ being a closed, but not exact, 1-form $p = dS$ on an SO(2)-reduction of GL(M). This amounts to saying that the flow defined by such a $p$ has zero *dynamical vorticity*:

$$\Omega \equiv dp = 0.$$

We now follow Takabayasi's lead and consider the generalization of the Madelung equations defined by letting $p$ be an arbitrary (but timelike) 1-form that satisfies the equations:

$$\delta p = 0,$$
$$p^2 = -\rho_0^2 c^2$$
$$\Box R - k_C^2 R[1 - R^4] = 0, \qquad (\rho_0 = m_0 R^2),$$

---

[9] Notice that we have succeeded in "decoupling" the equation for R from the equations for $p$.

but not the irrotationality constraint, so $\Omega = dp \neq 0$, in general.

Since we obtained the continuum-mechanical interpretation of wave mechanics by starting with the statistical interpretation and its Schrödinger-Klein-Gordon picture, this suggests that we are going beyond the scope of these equations. The question then becomes one of whether this expansion of scope has a corresponding statement in the former picture.

Takabayasi suggested that one way to introduce vorticity into the flow is to consider it to be a charged fluid with a charge density $\sigma$ that is moving in an electromagnetic field that is described by its Minkowski field strength 2-form F. In our notation, we first perform a minimal electromagnetic coupling of $p$ with the electromagnetic potential 1-form $\phi$:

$$p \rightarrow P = p + \tfrac{1}{c}\sigma\phi.$$

If P is irrotational and incompressible then we will have:

$$dp = -\tfrac{1}{c}\sigma F, \qquad \text{i.e.,} \qquad \Omega = -\tfrac{1}{c}\sigma F,$$
$$\delta p = -\tfrac{1}{c}\sigma\,\delta\phi.$$

In this form, the vorticity of $p$ is coupled to the electromagnetic field by means of the charge density.

If $\phi$ is given the usual Lorentz gauge $\delta\phi = 0$ then these equations can be consolidated into:

$$\Box p = \delta dp = -\tfrac{1}{c}\sigma\,\delta F = -\tfrac{1}{c}\sigma J,$$

in which J is the electromagnetic current 1-form. This last equation represents a forced wave equation for $p$ in which the forcing function is proportional to the charge density times the electromagnetic current.

More generally, this is also the way to couple an SO(2) (i.e., U(1)) gauge potential to the Klein-Gordon particle. The only thing that changes from the Klein-Gordon picture is the notion that the curvature (field strength) of the gauge potential couples to the dynamical vorticity of our extended particle's momentum density.

**7. Discussion.** In order to liberate quantum physics from the demoralizing tyranny of stochasticism and phenomenology, it is necessary to duplicate the successes of quantum physics by means of a set of first principles that do not rely on probability. This involves not only rederiving a lot of the established equations from more fundamental roots, but also rethinking the philosophy of physics that emerged from the Copenhagen school.

In the 1920's, when quantum mechanics was still "a-Born-ing," physics was at a sort of historical cusp. On the one hand, there was growing disenchantment with the former philosophy that everything in physics was explainable in terms of theoretical mechanics since the Michelson-Morley experiment seemed to contradict the assumption that electromagnetic waves propagated in the same manner as mechanical waves. (Indeed, continuum mechanics is rarely taught as a physics course nowadays, having been rusticated to the netherworld of engineering.) On the other hand, only the most enlightened physicists had accepted the geometrical picture of spacetime that Einstein's theory had introduced. In fact, Einstein was awarded the 1921 Nobel Prize in physics for his contributions to quantum theory, not his theory of relativity! Consequently, there was little interest in merely adapting continuum mechanics to the demands of a relativistic formulation, much less giving a continuum-mechanical formulation to a realm of phenomena where intuition seemed to fail even the best minds of science, namely, the realm of quantum phenomena.

At the present stage of physics, however, these former restrictions are not as acute. The inventory of experimentally studied quantum phenomena has been growing, along with the intuition that comes from seeing similar things happening in different circumstances. For instance, the appearance of spontaneous symmetry breaking and the role of fluctuations about classical equilibria or extremals seem to be common artifacts of quantum phenomena. At the same, considerably more is understood about the role of geometry and topology in the basic statements of mechanics and field theories. Furthermore, the appearance of complexity in nonlinear dynamical systems seems to suggest casting a new light upon old problems that were formerly approached by purely statistical methods – such as the theory of turbulence. Consequently, many of the original problems and theories of quantum mechanics take on a different character upon reassessment in light of current thinking.

The tentative conclusion to be drawn from the results described in this article is that relativistic continuum mechanics might still provide a viable basis for the first principles of quantum mechanics when one considers a few subtle adjustments to one's intuition about both subjects. First, the continuum in question should not be assumed to obey any of the common constitutive equations, which all have a distinctly macroscopic origin. In particular, the Madelung medium is not precisely a fluid. Second, one should concentrate primarily on the geometrical nature of continuum mechanics than the more conventional approaches. For instance, the deformations of the medium we are dealing with do not seem to take the form of diffeomorphisms of regions of spacetime so much as differentiable homotopies of the spacetime metric within the space of neighboring metrics. In particular, we showed that a conformal change of the Minkowski space metric induces a non-trivial connection whose curvature accounts for the difference between Newton's equation of motion and Schrödinger's equation, and whose holonomy integral implies the Bohr-Sommerfeld quantization rules. One might also consider the role of metric deformations generated by infinitesimal shears as well.

In subsequent research, the author intends to pursue the obvious problem of duplicating the usual consequences of quantum mechanics in its statistical form by means of analogous statements of a geometrical or topological nature, as well as the problem of reconstructing the statistical model as a corollary to the continuum model ([10]).

As far as the first problem is concerned, Takabayasi has already made some progress along those lines. For instance, the nonlinearity of the Madelung equations entails the consequence that the linear superposition of two single-particle wave functions, which would be a solution to the Schrödinger equations, corresponds to a continuum state with more complexity than a mere linear superposition. In particular, its dynamics seem to be more closely related to the Kepler problem, which leads to chaotic behavior. This might suggest a deeper nature to quantum uncertainty to be found in the complexity of nonlinear dynamical systems, not the errors in the measurement process. A reasonable approach to the second part of the problem, i.e., the reconstruction of the usual statistical interpretation, might be the approach of Misra and Prigogine [20].

---

[10] On the seventh day, he shall rest!

## Appendix: G-structures

A G-*structure* on a manifold M is a reduction [**21, 22**] of the bundle GL(M) of linear frames on M to a sub-bundle G(M) whose structure group G is a subgroup of GL(*n*). Such a sub-bundle consists of frames that are all contained in the same orbit of the action of G on all linear frames. To define a reduction is to choose a G-orbit in GL$_x$(M) at every point $x \in$ M. Any such orbit is diffeomorphic to GL(*n*)/G.

This means that we have defined a section of the associated fiber bundle whose principle bundle is GL(M) and whose fiber is GL(*n*)/G. In general, this bundle will not have a section, so the existence of a G reduction amounts to a problem in obstruction theory. However, in most cases the nature of that associated bundle is simpler than it looks.

For the reduction of GL(M) to O(M), the bundle has fibers that look like GL(*n*)/O(*n*), which is essentially the space of Riemannian metrics on $\mathbb{R}^n$. Hence, a section of the associated homogeneous bundle is simply a Riemannian structure *g* on M, which always exists for paracompact M. The reduction of GL(M) would consist of the bundle of orthonormal frames.

For a metric that is conformally related to g, the reduction would be a principal bundle isomorphic to the first, where the only difference would be in the choice of unit length for the vectors. In the case of a reduction to SO(M), one would also have to define a section of an associated O(*n*)/SO(*n*) = $\mathbb{Z}_2$ bundle, i.e., an orientation on M. A reduction of GL(M) to SO(*n*−1, 1)(M) amounts to a Lorentz structure on M. For non-compact M these always exist, but for compact M the obstruction is the vanishing of the Euler number of M [**23-24**].

Given a chain of subgroups of GL(*n*), GL(*n*) ⊃ $G_1$ ⊃ $G_2$ ⊃ …, if there is to be a reduction from a given subgroup to the next subgroup in the chain, there will have to be a section of the associated $G_i/G_{i+1}$-homogenous bundle and a corresponding obstruction class that relates to the existence of that section. For our present concerns, the issue in question is whether we can define a reduction of GL(M) to an SO(2) sub-bundle, at least for a four-dimensional M. A reasonable chain of reductions would be:

$$GL(4) \supset O(3,1) \supset SO(3,1) \supset SO(3) \supset SO(2).$$

Taking the first step means defining a Lorentz structure *g*, which reduces us to the bundle of Lorentz frames. The second step means choosing an orientation for our Lorentz frames. To reduce to oriented orthonormal 3-frames, all we need to do is fix one timelike vector at each point. This is simply a non-zero timelike vector field, so that involves the same obstruction as for time orientability of M. To reduce further to orthonormal 2-frames means fixing a spacelike vector at each point, i.e., defining a non-zero spacelike vector field. Here, we encounter a subtlety, since we have fixed two orthogonal non-zero vector fields, hence, a global orthonormal 2-frame field. For this to be possible, M must have degree of parallelizability equal to 2. The obstruction to this is the vanishing of the top two Stiefel-Whitney classes $w^3$(M), $w^4$(M) of T(M).